\documentclass[14pt,a4paper]{article}

\usepackage[english]{babel}
\usepackage{amsmath,amssymb,amsfonts,amsthm}
\usepackage[mathscr]{eucal}
\usepackage[all]{xy}
\usepackage{hyperref}
\usepackage{setspace}
\usepackage{upgreek}
\allowdisplaybreaks

\usepackage{times}
\usepackage{color}

\usepackage{indentfirst}

\begin{document}

\title{\textbf{Third order extensions  of $3d$
Chern-Simons interacting to gravity: Hamiltonian formalism and
stability}}
\author { \ D.~S.~Kaparulin\footnote{dsc@phys.tsu.ru},\ I.~Yu.~Karataeva\footnote{karin@phys.tsu.ru},\ and S.~L.~Lyakhovich\footnote{sll@phys.tsu.ru}}
\date{\footnotesize\textit{Physics Faculty, Tomsk State University, Tomsk 634050, Russia }}

\maketitle

\begin{abstract}
\noindent We consider inclusion of interactions between $3d$
Einstein gravity and the  third order extensions of Chern-Simons.
Once the gravity is minimally included into the third order vector
field equations, the theory is shown to admit a two-parameter series
of symmetric tensors with on-shell vanishing covariant divergence.
The canonical energy-momentum is included into the series. For a
certain range of the model parameters, the series include the
tensors that meet the weak energy condition, while the canonical
energy is unbounded in all the instances. Because of the on-shell
vanishing covariant divergence, any of these tensors can be
considered as an  appropriate candidate for the right hand side of
Einstein's equations. If the source differs from the canonical
energy momentum, the coupling is non-Lagrangian while the
interaction remains consistent with any of the tensors. We
reformulate these not necessarily Lagrangian third order equations
in the first order formalism which is covariant in the sense of
$1+2$ decomposition. After that, we find the Poisson bracket such
that the first order equations are Hamiltonian in all the instances,
be the original third order equations Lagrangian or not. The
brackets differ from canonical ones in the matter sector, while the
gravity admits the usual PB's in terms of ADM variables. The
Hamiltonian constraints generate lapse, shift and gauge
transformations of the vector field with respect to these Poisson
brackets. The Hamiltonian constraint, being the lapse generator, is
interpreted as strongly conserved energy. The matter contribution to
the Hamiltonian constraint corresponds to $00$-component of the
tensor included as a source in the right hand side of Einstein
equations. Once the $00$-component of the tensor is bounded, the
theory meets the usual sufficient condition of classical stability,
while the original field equations are of the third order.

\end{abstract}

\section{Introduction}
Various higher derivative field theories are studied once and again
over many decades for several reasons. Among the most frequently
mentioned advantages of the higher derivative systems are the better
convergence properties at classical and quantum level comparing to
the analogues without higher derivatives. For discussion of various
types of higher derivative models we refer to the paper \cite{KLS14}
and references therein. The higher derivative theories are also
notorious for the instability problem. The simplest stability test
-- boundedness of energy -- is usually failed by the models with
higher order equations of motion. The best known exception --
$f(R)$-gravity \cite{fR1} -- is stable due to very strong second
class constraints. Because of that, on the constrained surface, the
Hamiltonian is bounded, so the theory meets the sufficient condition
for stability.

Even if the energy is unbounded for general higher-derivative
dynamics, the theory is not necessarily unstable. If another bounded
conserved quantity exists, it stabilizes dynamics  at least at
classical level. For example, the free fourth-order Pais-Uhlenbeck
(PU) oscillator \cite{PU} performs bi-harmonic oscillations, being
obviously stable, while the canonical Ostrogradsky Hamiltonian is
unbounded in this model. The stability is due to the mere fact that
the sum of energies of two harmonics conserves, being the positive
quantity, while the canonical energy is a difference of energies of
the two oscillations. The true problem reveals itself once one
attempts to go beyond the free classical theory, either quantizing
the model or/and switching on the interaction at classical level. As
the canonical Hamiltonian formalism involves unbounded Ostrogradsky
Hamiltonian, being the difference of energies of harmonics, the
spectrum is unbounded. That results in instability at quantum level.
Various methods are discussed to cure the problem, see for example
\cite{Smilga1, BenM, Most}. These attempts are not systematically
extended beyond the free PU oscillator. It is also noticed that some
special interactions of PU model can have  the isles of stability in
classical dynamics, see \cite{Smilga2, Pavsic2, PUPert}, while it is
unstable globally.

In the paper \cite{KLS14}, a systematic way is suggested to switch
on the interactions in certain class of higher derivative systems
without breaking stability of free theory. This class of higher
derivative systems encompasses both PU oscillator with various
extensions, and a range of field theories. The class of higher
derivative dynamics with stable interactions has been further
extended in the paper \cite{KKL}. The starting point for this scheme
of switching on stable interactions is that the free higher
derivative theory should admit a series of conserved quantities such
that includes the canonical energy. Than the method implies to
switch on the interaction in such a way that is compatible with
conservation of certain representative of this series. Another
aspect of this scheme is that any of the conserved quantities of the
same series, not just canonical energy, is connected to same the
symmetry of the equations. The connection is established by the
extension of Noether theorem suggested in the paper \cite{KLS10}.
This implies that the field equations admit Lagrange anchor, which
has been first introduced in ref. \cite{KazLS} to BRST embed and
quantize not necessarily Lagrangian systems. The Lagrange anchor,
being admitted by the equations of motion, connects the conserved
quantities with symmetries \cite{KLS10} irrespectively to existence
of the action functional for the equations. One more crucial feature
of the Lagrange anchor is that it makes the theory Hamiltonian in
the first order formalism \cite{KLS-m}. This means, if the higher
derivative equations admit multiple Lagrange anchors, the first
order formulation is a multi-Hamiltonian theory. Given the Lagrange
anchor in free theory, the interaction can be switched in such a way
that provides conservation of the quantity connected by the anchor
with the symmetry if the latter is unbroken at interacting level
\cite{KLS16}. In particular, if the conserved quantity is bounded in
free theory, the dynamics will remain stable upon inclusion of
interaction, at least at perturbative level. One more detail of this
scheme of introducing the interactions is that the interacting
theory will admit constrained Hamiltonian formulation even if the
vertices are non-Lagrangian in the original higher derivative
equations.

The free PU oscillator provides a simplest example of a
multi-Hamiltonian higher derivative system, as it has been noticed
more than a decade ago, \cite{BK, DS}. The multi-Hamiltonian
formulations are also known for various extensions of the free PU
oscillator \cite{Mas15, Mas16}. It has been also found that the
interactions can be included into the PU equations leaving the
dynamics stable \cite{KLS14}. Furthermore, the stable higher
derivative PU equation with interaction still admits Hamiltonian
formulation in the corresponding first order formalism, though the
stable vertices are non-Lagrangian in the higher derivative
equations of motion \cite{KL15}.

One of the frequently discussed higher derivative field theories is
the third order extension of  Chern--Simons \cite{Deser}. It has
been recently found that the model is multi-Hamiltonian at free
level \cite{AKL18}. To the best of our knowledge, it is the first
known explicit example of higher derivative field theory that admits
multi-Hamiltonian structure. It is also found that the stable
interaction can be included with spin $1/2$ in this model
\cite{AKL18}. To keep dynamics stable, the explicitly covariant
interaction vertices should be non-minimal and non-Lagrangian
\cite{AKL18}. Even though the stable interaction is non-Lagrangian,
the theory still admits Hamiltonian formulation at interacting level
\cite{AKL18}, so it can be quantized.

In this paper, we study the coupling of the third order extension of
Chern-Simons to Einstein's gravity. At free level, the theory admits
a continuous series of conserved tensors found in \cite{KKL} that
includes canonical energy-momentum. The series involves bounded
quantities if the free third order field equations describe
reducible unitary representations, while the canonical
energy-momentum is unbounded in every instance.

We begin with inclusion of minimal coupling  to gravity into the
third order field equations
\begin{equation}\label{EoM}
(m^{-1}\ast d\ast d \ast d+\alpha_2\ast d\ast d+\alpha_1m\ast
d)A=0\,,
\end{equation}
where $A=A_\mu(x)dx^\mu, \, \mu=0,1,2$ is the vector field, $m$ is
the constant with the dimension of mass, $\alpha_1,\alpha_2$ are the
dimensionless constant parameters, $d$ denotes the exterior
derivative, and $\ast$ stands for the Hodge star operator,
\begin{equation}\label{d-ast}
   (\ast dA)^\mu=\frac{1}{\sqrt{g}}\varepsilon^{\mu\nu\rho}
    \partial_{\nu}A_{\rho}\,.
\end{equation}
All the tensor indices are raised and lowered by the spacetime
metric $g_{\mu\nu}$. The signature of the metric is mostly negative.
The $3d$ Levi-Civita symbol $\varepsilon^{\mu\nu\rho}$ is the tensor
density, $\varepsilon^{012}=1$.

If field equations (\ref{EoM}) are considered in Minkowski space,
they admit a series conserved tensors found in \cite{KKL}. As we
shall demonstrate, the minimal covariant extension of this series
leads us to the covariantly transverse tensors
\begin{equation}\label{T-b}
    \nabla_\nu T^{\mu\nu}(A,g; \alpha, \beta, \gamma)\approx 0 \, , \qquad
    \forall \alpha, \beta, \gamma.
\end{equation}
Here $\alpha$ are the parameters involved in the field equations
(\ref{EoM}), while $\beta, \gamma$ are the independent real
parameters labeling the representatives of the series of tensors.
The sign $\approx$ means the on-shell equality with respect to the
equations (\ref{EoM}). The explicit expressions for the series of
tensors $T$ are provided in the next section. Once the tensors
(\ref{T-b}) are on shell covariantly transverse, any representative
of the series can be considered as an admissible right hand side for
the Einstein equations\footnote{We adopt the following definitions
for covariant derivative $\nabla_\mu$, curvature tensor
$R^{\mu}{}_{\nu\rho\tau}$ and Ricci tensor $R_{\mu\nu}$:
$$
\nabla_\mu\, A_\nu = \partial_\mu\, A_\nu - \Gamma_{\, \mu\nu}{}^{\,
\rho} \, A_\rho\,,\qquad R^\rho{}_{\tau\mu\nu} \ =\
 \partial_\mu \, \Gamma_{\nu\tau}{}^{\, \rho} \ -\ \partial_\nu \, \Gamma_{\mu\tau}{}^{\, \rho} \ +\
 \Gamma_{\mu\sigma}{}^{\, \rho}\, \Gamma_{\nu\tau}{}^{\, \sigma} \ -\  \Gamma_{\nu\sigma}{}^{\, \rho}\, \Gamma_{\mu\tau}{}^{\, \sigma}\,,\qquad
R_{\mu\nu} = R^\rho{}_{\mu\rho\nu}\,,
$$
where  $\Gamma$ is the Christoffel symbol; $\Lambda$ is the
cosmological constant.}
\begin{equation}\label{ET}
    R^{\mu\nu}-\frac{1}{2}g^{\mu\nu}(R-\Lambda)=-\frac{1}{2}T^{\mu\nu}(A,g; \alpha, \beta, \gamma)
    \, .
\end{equation}
If the rhs of these equations is the canonical energy-momentum for
the field $A$ subject to the equations (\ref{EoM}), then the
equations (\ref{EoM}), (\ref{ET}) form a Lagrangian system.
Otherwise, it is not Lagrangian.  Be the equations Lagrangian or
not, the system (\ref{EoM}), (\ref{ET}) is  fully consistent once
the tensor $T$ on the rhs of (\ref{ET}) is transverse on shell
(\ref{EoM}).

There are two obvious facts indicating the consistency of field
equations (\ref{EoM}) and (\ref{ET}): (i) the system has explicit
gradient gauge symmetry for the field $A$, and the equations are
diffeomorphism invariant; (ii) there are gauge identities between
the equations which are the same in Lagrangian and non-Lagrangian
case -- the divergence vanishes identically of the equations
(\ref{EoM}), while the covariant divergence of equations (\ref{ET})
vanishes because of (\ref{T-b}). The orders of equations,
symmetries, and identities are the same in all the cases, Lagrangian
and non-Lagrangian. These data are sufficient to define the degree
of freedom number for the theory being formulated in covariant form,
without explicit recourse to Hamiltonian constrained analysis. The
formula (8) of the paper \cite{KLS-JHEP13} allows one to easily
compute the local degree of freedom number in a covariant way. The
computation gives that the number is four,\footnote{We mean the
usual definition for the local DoF number -- it is the number of
independent Cauchy data per point of space. } i.e. it is the same as
for the equations (\ref{EoM}) in Minkowski space. This means
consistency, because in three dimensions, Einstein's gravity does
not have local DoF by itself.

In the next section, we elaborate on the on-shell transverse tensors
admitted by the equations (\ref{EoM}). In particular, we use the ADM
space decomposition to clarify the structure of the tensors. As we
shall see, the series of tensors $T(\alpha,\beta,\gamma)$
(\ref{T-b}) includes the representatives  that meet the so-called
weak energy condition (abbreviated as WEC),  once the parameters
$\alpha, \beta$ meet certain conditions.\footnote{The accessory
parameter $\gamma$ has a different meaning. On shell, the tensors
$T(\alpha,\beta,\gamma)$ become $\gamma$-independent. So this
parameter labels representatives in the same equivalence class of
the on-shell transverse tensors. Given the parameters $\alpha$
defining the the field equations (\ref{EoM}), we have the two
parameters $\beta$ labeling inequivalent on-shell transverse
tensors. } The dynamics is stable once the tensor in the rhs of
Einstein equations (\ref{ET}) meets the WEC. The canonical
energy-momentum, being included in the series, does not meet the
condition, so the stable interactions are inevitably non-Lagrangian.
In the section three, we reformulate the field equations
(\ref{EoM}), (\ref{ET}) in the first order formalism with respect to
the time derivatives. This is done making use of $1+2$ decomposition
in the ADM variables. Then, we find the Poisson bracket such that
the first order equations read as a constrained Hamiltonian system
in all the instances, be the original system (\ref{EoM}), (\ref{ET})
Lagrangian or not. The Poisson bracket is not canonical in general,
and it involves the parameters $\alpha,\beta,\gamma$ from the rhs of
the Einstein's equations (\ref{ET}).  All the Hamiltonian
constraints are of the first class with respect to this bracket. The
constraints include the Hamiltonian generators of lapse and shift
transformations, and also the generator of gauge transformations for
the vector field. The matter contribution to the lapse constraint
can be positive for certain range of the model parameters
$\alpha,\beta$ involved in the field equations (\ref{EoM}),
(\ref{ET}). If the rhs of Einstein's equations (\ref{ET}) is the
canonical energy-momentum (that corresponds to $\beta_1=1, \,
\beta_2=0$ ), the matter contribution to the lapse constraint will
be unbounded for any $\alpha$. In this case, the constrained
Hamiltonian formalism is canonically equivalent to Ostrogradsky
formulation of the Lagrangian theory. The formulations with the
bounded Hamiltonian are inequivalent to this case, because no
canonical transformation can turn any on-shell bounded quantity into
an unbounded one and vice versa.

\section{Tensors with on-shell vanishing covariant
divergence and stability}

In this section, we find the series of on-shell covariantly
transverse second rank symmetric tensors, and study the weak energy
condition for these tensors.

Let us introduce abbreviations
\begin{equation}\label{FG-notation}
    F=\ast dA\,,\qquad G=\ast d\ast d A\,.
\end{equation}
Obviously, $F$ and $G$ are gauge invariant quantities. Also notice
that from the definition (\ref{FG-notation}) immediately follows
that the one-forms $F,G$ are co-closed, so the covariant divergence
identically vanishes of the vectors $F^\mu,G^\mu$,
\begin{equation}\label{div-GF}
d*F\equiv 0 \, , \quad d*G\equiv 0 \,  \qquad \Leftrightarrow \qquad
\nabla_\mu F^\mu\equiv 0 \, , \quad \nabla_\mu G^\mu\equiv 0 \, .
\end{equation}
 In terms of $F,G$, the third order equations (\ref{EoM}) read
\begin{equation}\label{Egamma}
    E\equiv m^{-1}\ast dG+\alpha_2 G+\alpha_1m F\approx 0\qquad \Leftrightarrow \qquad
    E^{\mu}\equiv m^{-1}\frac{1}{\sqrt{g}}\varepsilon^{\mu\nu\rho}\partial_\nu G_\rho+\alpha_2G^\mu+\alpha_1mF^\mu\approx 0\,.
\end{equation}
In the paper \cite{KKL}, two independent on-shell conserved
symmetric tensors are found for these equations in Minkowski space.
The minimal covariant extension of these tensors\footnote{In ref.
\cite{KKL} the Chern-Simons extension of $n$-th order is considered
in the flat space, and the series of conserved tensors is
$n-1$-parametric. Here, we consider the coupling to gravity only for
the third order extension.} read
\begin{align}\label{T1}
    T_{(1)}^{\mu\nu}(A,g;\alpha)&=\frac{1}{m}\Big[(G^\mu F^\nu+G^\nu
    F^\mu-g^{\mu\nu}G_\rho
    F^\rho)+\alpha_2m(F^\mu
    F^\nu-\frac{1}{2}g^{\mu\nu}F_\rho F^\rho)\Big]\,,\\[3mm]\label{T2}
    T_{(2)}^{\mu\nu}(A,g;\alpha)&=\frac{1}{m^2}\Big[(G^\mu
    G^\nu-\frac{1}{2}g^{\mu\nu}G_\rho
    G^\rho)-\alpha_1m^2(F^\mu
    F^\nu-\frac{1}{2}g^{\mu\nu}F_\rho F^\rho)\Big]\,,
\end{align}
where $\alpha_1,\alpha_2$ are the parameters of the third-order
extension of Chern-Simons (\ref{EoM}). Upon account for the
identities (\ref{div-GF}), the covariant divergence of the tensors
(\ref{T1}), (\ref{T2}) is seen to vanish on shell:
\begin{equation}\label{div-CT-series}\begin{array}{l}\displaystyle
    \nabla_\nu
    T_{(1)}^{\mu\nu}(A,g;\alpha)=-\frac{1}{\sqrt{g}}\varepsilon^{\mu\nu\rho}F_\nu E_\rho\approx
    0\,,\qquad
    \nabla_\nu
    T_{(2)}^{\mu\nu}(A,g;\alpha)=-\frac1m\frac{1}{\sqrt{g}}\varepsilon^{\mu\nu\rho}G_\nu E_\rho\approx
    0\,.
\end{array}\end{equation}
Notice that any on-shell vanishing tensor is on-shell transverse. In
the third-order Chern-Simons theory (\ref{EoM}), one of these
trivial tensors is relevant for constructing Hamiltonian
formulation. We chose it in the form
\begin{equation}\label{T3}\begin{array}{rl}\displaystyle
    T^{\mu\nu}_{(3)}(A,g;\alpha,\beta,\gamma)&=\displaystyle\frac{1}{m^2}\displaystyle\frac{\beta_2^2+\gamma\beta_1}{\beta_1-\beta_2\alpha_2-\gamma\alpha_1}
    \Big(E^\mu G^\nu+E^\nu G^\mu-g^{\mu\nu}E_\rho G^\rho\Big)+\\[3mm]\displaystyle&\displaystyle
    +\frac{1}{m}\frac{\beta_1\beta_2+\gamma(\beta_1\alpha_2-\beta_2\alpha_1)}{\beta_1-\beta_2\alpha_2-\gamma\alpha_1}
    \Big(E^\mu F^\nu+E^\nu F^\mu-g^{\mu\nu}E_\rho
    F^\rho\Big)\,,
\end{array}\end{equation}
where $E^\mu$ denote the lhs of equations (\ref{Egamma}), and
$\beta_1,\beta_2,\gamma$ are constant dimensionless parameters such
that $\beta_1-\beta_2\alpha_2-\gamma\alpha_1\neq0$.

As the tensors (\ref{T1}), (\ref{T2}), (\ref{T3}) have on-shell
vanishing covariant divergence, any linear combination of these
tensors is covariantly transverse on shell. Given the field
equations (\ref{EoM}), we have the three-parameter series of
on-shell transverse tensors
\begin{align}\notag
    &T^{\mu\nu}(A,g;\alpha,\beta,\gamma)=\beta_1
    T_{(1)}^{\mu\nu}(A,g;\alpha)+\beta_2T_{(2)}^{\mu\nu}(A,g;\alpha)+T^{\mu\nu}_{(3)}(A,g;\alpha,\beta,\gamma)=\\[4mm]\notag
    &=\frac{\beta_1}{m}\Big[(G^\mu F^\nu+G^\nu F^\mu-g^{\mu\nu}G_\rho
    F^\rho)+\alpha_2m(F^\mu F^\nu-\frac{1}{2}g^{\mu\nu}F_\rho
    F^\rho)\Big]+\\[3mm]\notag&
    +\frac{\beta_2}{m^2}\Big[(G^\mu G^\nu-\frac{1}{2}g^{\mu\nu}G_\rho
    F^\rho)-\alpha_1m^2(F^\mu F^\nu-\frac{1}{2}g^{\mu\nu}F_\rho
    F^\rho)\Big]+\\[3mm]&+\frac{1}{m^2}
    \frac{\beta_2^2+\gamma\beta_1}{\beta_1-\beta_2\alpha_2-\gamma\alpha_1}
    \Big(E^\mu G^\nu+E^\nu G^\mu-g^{\mu\nu}E_\rho G^\rho\Big)+
    \frac{1}{m}\frac{\beta_1\beta_2+\gamma(\beta_1\alpha_2-\beta_2\alpha_1)}{\beta_1-\beta_2\alpha_2-\gamma\alpha_1}\Big(E^\mu F^\nu+E^\nu F^\mu-g^{\mu\nu}E_\rho
    F^\rho\Big)\Big]\,,
\label{CT-series}\end{align} where $\beta_1,\beta_2, \gamma$ are the
independent parameters that label representatives of the series,
while $\alpha_1,\alpha_2$ specify the field equations. Once the
tensor $T^{\mu\nu}(A,g;\alpha,\beta,\gamma)$ is included into rhs of
Einstein's equations (\ref{ET}), $\beta_1,\beta_2$ being the factors
at the on-shell non-vanishing contributions (\ref{T1}), (\ref{T2}),
can be understood as coupling constants responsible for the
interaction between gravity and matter. The accessory parameter
$\gamma$ is involved into on-shell vanishing term. So, it accounts
for possible contributions to the rhs of equations (\ref{ET}) such
that vanish on account of  the field equations (\ref{EoM}).

The equations (\ref{EoM}) for the vector field, being considered
alone, apart from Einstein's equations (\ref{ET}), follow from the
least action principle for the functional
\begin{equation}\label{S-A}
S_A=\frac{1}{2}\int \ast A\wedge(m^{-1}\ast d\ast d\ast
d+\alpha_2\ast d\ast d+\alpha_1m\ast d)A\, , \qquad \frac{\delta
S_A}{\delta A_\mu}\equiv\sqrt{g} E^\mu \, .
\end{equation}
The  quantity $T_{(1)}^{\mu\nu}(A,g;\alpha)$ (\ref{T1}), being the
first constituent of the series of on-shell transverse tensors
(\ref{CT-series}), is the canonical stress-energy tensor for the
action,
\begin{equation}\label{S-E}
T_{(1)}^{\mu\nu}(A,g;\alpha) \equiv \frac{2}{\sqrt{g}}\frac{\delta
S_A}{\delta g_{\mu\nu} }\,.
\end{equation}
while $T_{(2)}^{\mu\nu}(A,g;\alpha)$ (\ref{T2}) is a different
independent tensor. If the scalar curvature of the metric and
cosmological constant are added to the matter action (\ref{S-A}),
the corresponding Lagrange equations will be Einstein's ones
(\ref{ET}) with canonical stress-energy tensor
$T_{(1)}^{\mu\nu}(A,g;\alpha)$ (\ref{T1}) in the right hand side.
Contribution of $T_{(2)}^{\mu\nu}(A,g;\alpha)$ (\ref{T2}) into the
right hand side of equations (\ref{ET}) is on-shell non-trivial.
This contribution is non-Lagrangian, though it is fully consistent
as it has been already explained in the introduction.

Below we elaborate on the issue of stability of the system
(\ref{EoM}), (\ref{ET}). As we shall see, the stability can be
achieved with certain representatives of the series
(\ref{CT-series}) in the right hand side of Einstein's equations,
while inclusion of a pure canonical stress-energy results in
instability.

Various assumptions are known about energy-momentum tensor which can
provide stability of dynamics of gravity coupled to the matter.
These assumptions are usually referred to as energy conditions. For
general discussion of energy conditions we refer to the books
\cite{Wald, Caroll}. In this paper, we examine the week energy
condition (WEC), which implies that the scalar
\begin{equation}\label{WEC}
    T^{\mu\nu}(A,g;\alpha,\beta,\gamma)\xi_\mu\xi_\nu\geq0\,,
\end{equation}
is bounded  from below on the mass shell (\ref{Egamma}) for
arbitrary timelike vector $\xi_\mu(x), \, \xi^2 >0 $. This condition
means that the observer will always measure a positive energy
density of the matter when traveling by any timelike path. Since the
Hamiltonian of the theory is the phase space equivalent of energy,
the theory, whose matter satisfies the WEC (\ref{WEC}), has a good
chance to admit a Hamiltonian formulation with bounded Hamiltonian
of the matter. The latter can be viewed as the stability condition
in the usual sense.

Once the WEC (\ref{WEC})  is imposed onto the tensor
$T(\alpha,\beta,\gamma)$ (\ref{CT-series}), it restricts the
admissible range of parameters $\alpha,\beta$. Now, we are going to
find these restrictions explicitly. Let us choose a special
coordinate system such that the timelike vector $\xi$ has the
canonical form $\xi_\mu=(1,0,0)$. In this coordinate system, the
inequality (\ref{WEC}) reads
\begin{equation}\label{WEC-00}
    T^{00}(A,g;\alpha,\beta,\gamma)\geq0\,.
\end{equation}
The lhs of this expression is a bilinear form in the variables
$F,G$. The coefficients of the form depend on the metric. To
simplify the dependence on metric, we use the ADM variables that
suites well to the problems where the explicit decomposition in
space and time has to be done. The ADM variables read
\begin{equation}\label{N-gast}
N \equiv\frac{1}{\sqrt{g^{00}}}\,,\qquad N_i \equiv g_{0i}\,,\qquad
\buildrel\ast\over{g}_{ij} \equiv g_{ij}\,, \qquad i,j=1,2\,.
\end{equation}
The $3d$ metric  $g_{\alpha\beta}$ and its inverse can be expressed
in terms of the ADM variables,
\begin{equation}\label{ADM-var}
 g_{\alpha\beta}=
\left(
  \begin{array}{cc}
    N^2  + N_s N^s  \ \ & N_j \\
    N_i\ \  & \buildrel\ast\over{g}_{ij}  \\
  \end{array}
\right)\,,
 \qquad
 g^{\alpha\beta}=
\left(
  \begin{array}{cc}
    N^{-2} \ \  & - N^j N^{-2} \\
   - N^i N^{-2}\ \  & \buildrel\ast\over{g}{\!}^{ij} + N^i N^j N^{-2}\\
  \end{array}
\right)\,,
\end{equation}
\begin{equation}\label{g-ast}
    \buildrel\ast\over{g}{\!}^{is} \buildrel\ast\over{g}_{sj} = \delta^i_j\,,\qquad N^i\equiv\,\buildrel\ast\over{g}{\!}^{ij}
    N_j\,.
\end{equation}
For the vector fields $E$ (\ref{Egamma}), $F\,,G$
(\ref{FG-notation}), we introduce the following $1+2$ decomposition:
\begin{equation}\label{Ftilde}
Z_\alpha=(N\widehat{Z}^0+N^sZ_s,Z_i)\,,\qquad \widehat{Z}^0=N
Z^0\,,\qquad Z=\{E,G,F\}\,,
\end{equation} where $E_i\,,G_i\,,F_i$ are space
components of the forms (\ref{Egamma}), (\ref{FG-notation}), and
$E^0\,,G^0\,,F^0$ are the time components of the $3d$ vectors
$F^\alpha,G^\alpha$. The quantities
$\widehat{E}^0\,,\widehat{G}^0\,,\widehat{F}^0$ read
\begin{equation}\label{Etilde}
\theta\equiv\widehat{E}^0=\frac{1}{\sqrt{\buildrel\ast\over{g}}}\varepsilon^{sr}
(m^{-1}\partial_sG_r+\alpha_2\partial_sF_r+\alpha_1m\partial_sA_r)\,,\qquad
\widehat{G}^0=\frac{1}{\sqrt{\buildrel\ast\over{g}}}\varepsilon^{sr}\partial_{s}F_r\,,\qquad
\widehat{F}^0=\frac{1}{\sqrt{\buildrel\ast\over{g}}}\varepsilon^{sr}\partial_{s}A_r\,,
\end{equation} where $\varepsilon^{12}=1$. Here, the
mass shell condition $\theta=0$ corresponds to the Gauss law
constraint in the model (\ref{EoM}).

In the notation (\ref{Ftilde}), (\ref{Etilde}), we rewrite the WEC
(\ref{WEC}) in the ADM variables
\begin{equation}\nonumber
T^{00} = \frac{1}{N^2} \frac{\beta_1}{m}\bigg[ \big( \widehat{G}^0
\widehat{F}^0   - \buildrel\ast\over{g}{\!}^{sr} {G}_s {F}_r \big) +
\alpha_2 m \frac{1}{2} \big( \widehat{F}^{0} \widehat{F}^0  -
\buildrel\ast\over{g}{\!}^{sr} {F}_s {F}_r \big) \bigg] +
\end{equation}
\begin{equation}\nonumber
+ \frac{1}{N^2} \frac{\beta_2}{m^2}\bigg[ \frac{1}{2} \big(
\widehat{G}^0 \widehat{G}^0  -  \buildrel\ast\over{g}{\!}^{sr} {G}_s
{G}_r \big) - \alpha_1 m^2 \frac{1}{2} \big( \widehat{F}^{0}
\widehat{F}^0 - \buildrel\ast\over{g}{\!}^{sr} {F}_s {F}_r \big)
\bigg] +
\end{equation}
\begin{equation}\label{T00-series-WEC}
+\ \frac{1}{N^2} \frac{1}{m^2}\ \frac{\beta_2^2   + \gamma\
\beta_1}{\beta_1 - \beta_2\alpha_2 - \gamma\alpha_1} \big(
\widehat{G}^0  \theta - \buildrel\ast\over{g}{\!}^{sr} {E}_s{G}_r
\big)  + \frac{1}{N^2} \frac{1}{m}\ \frac{\beta_1 \beta_2 + \gamma
(\beta_1\alpha_2 - \beta_2\alpha_1)}{\beta_1 - \beta_2\alpha_2 -
\gamma\alpha_1} \big(\widehat{F}^0 \theta-
\buildrel\ast\over{g}{\!}^{sr} {E}_s{F}_r \big)\geq0\,.
\end{equation}
We evaluate the lhs of this inequality on the mass shell
(\ref{Etilde}), being equivalent to the original equations
(\ref{EoM}). On shell, \mbox{$\theta=0,\,E_i=0$}, while the rest of
the expression is a bilinear form in the variables $G_i,F_i,A_i$.
Only $4$ of these variables give independent Cauchy data for the
model (\ref{EoM}): one of components $A_i$ can be set to zero by
gradient gauge transformation, and one of components $G_i$ is fixed
by the Gauss law constraint $\theta=0$. Choosing initial data in the
form
\begin{equation}\label{Fxizeta}
G_i=m\partial_i\xi\,,\qquad F_i=\partial_i\zeta\,,\qquad A_i=0\,,
\end{equation} with
$\xi(x),\zeta(x)$ being the arbitrary functions of spacetime
coordinates, we automatically satisfy the Gauss law constraint,
while the condition (\ref{T00-series-WEC}) reads
\begin{equation}\label{T00-xi-eps}
    -\frac{1}{2}\beta_1{\!}\buildrel\ast\over{g}{\!}^{ij}\partial_i\xi
    \partial_j\xi+\beta_2{\!}\buildrel\ast\over{g}{\!}^{ij}\partial_i\xi
    \partial_j\zeta-\frac{1}{2}(\beta_1\alpha_2-\beta_2\alpha_1)\buildrel\ast\over{g}{\!}^{ij}\partial_i\zeta\partial_j\zeta\geq0\,,\qquad
    \forall\zeta(x),\xi(x)\,.
\end{equation}
Considering this expression as the quadratic form in the gradient
vectors $\partial_i\xi,\partial_i\zeta$, and using the Sylvester
criterion to ensure that the form is positive semidefinite, we get
two restrictions on the parameters $\alpha,\beta$,
\begin{equation}\label{Stability-condidtions}
\beta_2\geq0\,,\qquad
-\beta_1^2+\beta_1\beta_2\alpha_2-\beta_2^2\alpha_1\geq0\,.
\end{equation}
These two conditions are also sufficient to meet WEC (\ref{WEC}),
because, in this case, (\ref{T00-series-WEC}) is a positive
(semi-)definite quadratic form of the variables
$\widehat{G}^0,\widehat{F}^0,G_i,F_i$. \footnote{As we use mostly
negative signature of metric in 1+2 dimensions, the quadratic form
like $(k^0)^2 -{\stackrel{\ast}{g}}{}^{ij}k_ik_j$ is positive for
any vector $k_\mu$.} There is a special case when the quadratic form
$T^{00}(\alpha,\beta,\gamma)$ (\ref{T00-series-WEC}) is degenerate.
We skip the degenerate case in this paper. The non-degeneracy
requirement restricts the parameters by the condition
\begin{equation}\label{b3-neq-0}
    \beta_1^2-\beta_1\beta_2\alpha_2+\beta_2^2\alpha_1\neq0\,.
\end{equation}
This restriction is assumed in all the considerations below.

Relations (\ref{Stability-condidtions}), (\ref{b3-neq-0}) determine
the range of the parameters $\alpha,\beta$ that results in the
stable theory (\ref{EoM}), (\ref{ET}). The consistency for these
relations implies that the parameters $\alpha$ have to meet the
condition
\begin{equation}\label{ab}
    \alpha_2^2-4\alpha_1>0\,.
\end{equation}
Under this condition, the Minkowski space limit of the third order
Chern-Simons equations (\ref{EoM}) transforms by reducible unitary
representation of the Poincar\'e group, see ref. \cite{KKL}. If
$\alpha_1\neq0$, it describes the pair of self-dual massive spins 1
 with different masses. The self-dual massive spin models are well known in $3d$ \cite{Townsend,DJ1984}.
 If $\alpha_1=0$, the set of sub-representations includes a massless spin 1 and a massive
spin 1 subject to a self-duality condition. Once relation (\ref{ab})
is not satisfied, the Minkowski space limit of the equations
(\ref{EoM}) transforms by a non-unitary representation of the
Poincar\'e group. The non-unitary representation does not decompose
into irreducible ones. In this case, any tensor of the series
(\ref{CT-series}) does not meet the WEC (\ref{WEC}), so the dynamics
is unstable with any $T$ of the series.

The general conclusion is that the model (\ref{EoM}) with any
parameters $\alpha_1, \alpha_2$ admits a series of consistent
couplings with gravity described by tensors (\ref{div-CT-series})
included in the rhs of Einstein's equations (\ref{ET}). If the WEC
(\ref{WEC}) is imposed, it ensures the stability of dynamics. From
the WEC, the restriction follow on the parameters
$\alpha_1,\alpha_2$ of the extended CS equations (\ref{EoM}). Given
the parameters $\alpha_1,\alpha_2$, the parameters $\beta_1,
\beta_2$ define the admissible tensor (\ref{CT-series}) in the rhs
of Einstein's equations (\ref{ET}). If the WEC is imposed, the
parameters $\beta$ have to meet the conditions
(\ref{Stability-condidtions}). Under these conditions the third
order theory (\ref{EoM}) remains stable being coupled to Einstein's
gravity.

\section{The first order formulation}
The existence of Hamiltonian formulation of the model turns out
indifferent to stability. So, we develop the first order formalism
for equations (\ref{EoM}), (\ref{ET}) and seek for the Hamiltonian
structure in a uniform way for any tensor of the series
(\ref{CT-series}) included in the rhs of equations (\ref{ET}), be
the model stable or unstable. Let us notice once again, that the
field equations (\ref{EoM}), (\ref{ET}) are non-Lagrangian in all
the instances besides $\beta_2=0$ which results in the unstable
theory, be the flat space limit stable or not.

When the Hamiltonian formulation is constructed for the
diffeomorphism-invariant theories, it is convenient to represent the
spacetime as a foliation whose leaves are the spacelike
hypersurfaces.  Locally, the spacetime is understood as a normal
bundle to the spacelike hypersurface which is referred to as space.
The coordinate on the fiber of normal bundle is considered as time.
We suppose that we have the coordinates $x^\mu, \,\mu=0,1,2$ such
that $x^i,\,  i=1,2$ are the space coordinates, while $x^0$ is the
time. The ADM variables (\ref{N-gast}), (\ref{ADM-var}),
(\ref{g-ast}) are very convenient to describe the metric once the
spacetime is decomposed in space and time.

Let us specify the notation related to the ADM parametrization of
metric and curvature. In this section,
$\buildrel\ast\over{\nabla}_j$ denotes the covariant derivative with
respect to the space metric $\buildrel\ast\over{g}{\!}_{ij}$, all
the space indices are lowered and raised by
$\buildrel\ast\over{g}{\!}_{ij}$ and
$\buildrel\ast\over{g}{\!}^{ij}$. The scalar curvature of the space
metric $\buildrel\ast\over{g}{\!}_{ij}$ is $\buildrel\ast\over{R}$.

Once the derivatives of metric are to be considered in terms of
decomposition in space and time, the
 $2d$ tensor of extrinsic curvature of the spacelike hypersurface
$x^0=\text{const}$ is a relevant structure. It reads
\begin{equation}\label{K}
K_{ij} = \frac{1}{2N}\, \big[ \, \buildrel\ast\over{\nabla}_i N_j\
+\ \buildrel\ast\over{\nabla}_j N_i\ -\
\dot{\buildrel\ast\over{g}}{}_{ij}\, \big]\,.
\end{equation}
Hereinafter, the dot denotes derivative by $x^0$.  To absorb the
time derivatives of metric, we use the variable $\pi^{ij}$, which is
canonically conjugate momentum to $g_{ij}$ in Einstein's gravity
without matter. As the matter contribution to the Einstein's
equations (\ref{ET}), being expressed in terms of $F,G$
(\ref{FG-notation}), does not involve derivatives of metric, we
expect that the  inclusion of matter does not change the canonical
momentum of gravity. In terms of extrinsic curvature, the momentum
reads
\begin{equation}\label{pi}
\pi^{ij}=\sqrt{\buildrel\ast\over{g}}\,(\,\buildrel\ast\over{g}{\!}^{is}\buildrel\ast\over{g}{\!}^{jr}-
\buildrel\ast\over{g}{\!}^{ij}\buildrel\ast\over{g}{\!}^{sr}\,)\,K_{sr}
\,\qquad \Leftrightarrow\qquad
K_{ij}=\frac{1}{\sqrt{\buildrel\ast\over{g}}}\,(\,\buildrel\ast\over{g}{\!}_{is}\buildrel\ast\over{g}{\!}_{jr}-
\buildrel\ast\over{g}{\!}_{ij}\buildrel\ast\over{g}{\!}_{sr}\,)\,\pi^{sr}\,.
\end{equation}
The trace of $\pi^{ij}$ is denoted by
$\pi\equiv\,\buildrel\ast\over{g}{}_{sr}\pi^{sr}$. As is seen from
the definition, $\pi^{ij}$ and its trace $\pi$ are correspondingly
$2d$ tensor and scalar densities.

To depress the order of the field equations (\ref{EoM}), we
introduce the variables $F_i$, $G_i, i=1,2$ absorbing the first and
second time derivatives of the vector field $A$:
\begin{align}\label{F}
    F_i=&\frac{1}{N\sqrt{\buildrel\ast\over{g}}}\Big(\buildrel\ast\over{g}{\!}_{is}\,\varepsilon^{sr}(\partial_rA_0-\dot{A}_r)+N_i\,\varepsilon^{sr}\partial_sA_r\Big)\,,
    \\\label{G}
    G_i=&\frac{1}{N\sqrt{\buildrel\ast\over{g}}}\Big\{\buildrel\ast\over{g}{\!}_{is}\,\varepsilon^{sr}
    \Big(\partial_r\Big(N\frac{1}{\sqrt{\buildrel\ast\over{g}}}\,\varepsilon^{kl}\partial_kF_l+N^kF_k\Big)-\dot{F}_r\Big)+N_i\,\varepsilon^{sr}\partial_sG_r\Big\}\,.
\end{align}
The variables $F_i,G_i$ are the space components of the differential
forms $F,G$ (\ref{FG-notation}) in three dimensions, i.e. these can
be viewed as the reduction of the spacetime forms to the
hypersurface $x^0=const$.

In terms of these variables, the field equations (\ref{EoM}),
(\ref{ET}) read as the first order system:
\begin{align}\label{dA}
        & \dot{A}_i = N \sqrt{\buildrel\ast\over{g}}\,\, \varepsilon_{is}\, \buildrel\ast\over{g}{\!}^{sr} F_r +
    N^k \varepsilon_{ki}\,\, \varepsilon^{sr}\, \partial_s A_r  +  \partial_i  A_0\,;\\[5mm]\label{dF}
        & \dot{F}_i = N \sqrt{\buildrel\ast\over{g}}\,\, \varepsilon_{is}\, \buildrel\ast\over{g}{\!}^{sr} G_r +
    N^k \varepsilon_{ki}\,\,  \varepsilon^{sr}\, \partial_s F_r  +
    \partial_i \bigg( N\, \frac{1}{\sqrt{ \buildrel\ast\over{g}}}\,\, \varepsilon^{sr}\, \partial_{s}  A_{r} +  N^s  F_s \bigg)\,;\\[2mm]\notag
        & \dot{G}_i = - \alpha_2 m  N \sqrt{ \buildrel\ast\over{g}}\, \varepsilon_{is} \buildrel\ast\over{g}{\!}^{sr} G_r -
    \alpha_1 m^2  N \sqrt{ \buildrel\ast\over{g}}\, \varepsilon_{is}
    \buildrel\ast\over{g}{\!}^{sr} F_r - \alpha_2 m   N^k
    \varepsilon_{ki}\,\, \varepsilon^{sr}\, \partial_s F_r  -
    \alpha_1 m^2  N^k \varepsilon_{ki}\,\, \varepsilon^{sr}\, \partial_s A_r  +
    \\[3mm]\label{dG}&+ \partial_i  \bigg( N\, \frac{1}{\sqrt{ \buildrel\ast\over{g}}}\,\, \varepsilon^{sr}\, \partial_{s} F_{r} + N^s  G_s\bigg)\,;\\[2mm]\label{Gauss}
        & \theta \equiv\,\frac{1}{\sqrt{\buildrel\ast\over{g}}}\varepsilon^{sr}(m^{-1}\partial_s G_r+\alpha_2\partial_s F_r+\alpha_1  m\partial_s A_r)=0\,;\\[2mm]\label{dg}
        & \dot{\buildrel\ast\over{g}}{}_{ij} =\, \buildrel\ast\over{\nabla}_i N_j\ +\ \buildrel\ast\over{\nabla}_j N_i\ -\ 2N\,
    \frac{1}{\sqrt{\buildrel\ast\over{g}}}(\pi_{ij}-\buildrel\ast\over{g}{\!}_{ij}\pi)\,;\\[2mm]\notag
        & \dot\pi^{ij}=
    - \pi^{is} \buildrel\ast\over{\nabla}_s N^j -
    \pi^{js} \buildrel\ast\over{\nabla}_s N^i   +
    \sqrt{\buildrel\ast\over{g}} \buildrel\ast\over{\nabla}_s
    \Big(\frac{1}{\sqrt{\buildrel\ast\over{g}}} \pi^{ij} N^s\Big)
    +\\\notag&
    + \frac{1}{2} N \frac{1}{\sqrt{\buildrel\ast\over{g}}}
    \buildrel\ast\over{g}{}^{ij}\ ( \pi_{sr} \pi^{sr}- \pi^2 ) +
    \sqrt{\buildrel\ast\over{g}}\ \bigg[  \buildrel\ast\over{\nabla}{}^i
    \buildrel\ast\over{\nabla}{}^j  N - \buildrel\ast\over{g}{}^{ij}
    \buildrel\ast\over{\nabla}_s \buildrel\ast\over{\nabla}{}^s N \bigg]
    -   \frac{1}{2} N \sqrt{\buildrel\ast\over{g}}
    \buildrel\ast\over{g}{}^{ij}  \Lambda   -\\\notag& - \frac{1}{2}
    \frac{\beta_1}{m}N \sqrt{\buildrel\ast\over{g}}\,
    \buildrel\ast\over{g}{\!}^{ik}\buildrel\ast\over{g}{\!}^{jl} \bigg[
    \bigg( G_k F_l  + G_l F_k  - \buildrel\ast\over{g}_{kl} (
    \widehat{G}^0 \widehat{F}^0 +  \buildrel\ast\over{g}{\!}^{sr} G_s\,
    F_r ) \bigg)  + \alpha_2 m \bigg( F_k F_l  - \frac{1}{2}
    \buildrel\ast\over{g}_{kl}
    (\widehat{F}^0 \widehat{F}^0 +  \buildrel\ast\over{g}{\!}^{sr} F_s F_r) \bigg)
    \bigg] - \\[3mm]\label{dK}& - \frac{1}{2}  \frac{\beta_2}{m^2} N
    \sqrt{\buildrel\ast\over{g}}\,
    \buildrel\ast\over{g}{\!}^{ik}\buildrel\ast\over{g}{\!}^{jl} \bigg[
    \bigg( G_k G_l  - \frac{1}{2} \buildrel\ast\over{g}_{kl} (
    \widehat{G}^0 \widehat{G}^0 +  \buildrel\ast\over{g}{\!}^{sr} G_s
    G_r )\bigg) - \alpha_1 m^2 \bigg( F_k F_l - \frac{1}{2}
    \buildrel\ast\over{g}_{kl} (\widehat{F}^0 \widehat{F}^0 +
    \buildrel\ast\over{g}{\!}^{sr} F_s F_r) \bigg) \bigg]\,;
    \\[5mm]\notag
        & \tau \equiv\  \frac{1}{\buildrel\ast\over{g}}(\pi^2-\pi_{sr}\pi^{sr})-\buildrel\ast\over{R} +\Lambda+\\[1mm]\notag&+
    \frac{\beta_1}{m}\bigg[ \big( \widehat{G}^0 \widehat{F}^0 -
    \buildrel\ast\over{g}{}^{sr} {G}_s {F}_r \big)  + \alpha_2 m
    \frac{1}{2} \big( \widehat{F}^{0} \widehat{F}^0  -
    \buildrel\ast\over{g}{}^{sr} {F}_s {F}_r \big) \bigg]\ +
    \\[3mm]\notag& +\frac{\beta_2}{m^2}\bigg[ \frac{1}{2} \big(
    \widehat{G}^0 \widehat{G}^0  -  \buildrel\ast\over{g}{}^{sr}
    {G}_s
    {G}_r \big) - \alpha_1 m^2 \frac{1}{2} \big( \widehat{F}^{0}
    \widehat{F}^0 - \buildrel\ast\over{g}{}^{sr} {F}_s {F}_r \big)
    \bigg]+\\[3mm]\label{T}&+
    \frac{1}{m^2}\frac{\beta_2^2+\gamma\beta_1}{\beta_1-\beta_2\alpha_2-\gamma\alpha_1}
    \widehat{G}^0\theta+\frac{1}{m}\frac{\beta_1\beta_2+\gamma(\beta_1\alpha_2-\beta_2\alpha_1)}
    {\beta_1-\beta_2\alpha_2-\gamma\alpha_1}\widehat{F}^0\theta=0\,;\\[5mm]\notag
        & \tau_i \equiv  -2\buildrel\ast\over{\nabla}_s\,
    \bigg(\frac{1}{\sqrt{\buildrel\ast\over{g}}}\pi^{s}{}_i\bigg)+
    \frac{\beta_1}{m}\bigg[ \big( \widehat{G}^0 {F}_i  + {G}_i
    \widehat{F}^0 \big)  + \alpha_2 m \widehat{F}^0 {F}_i
    \bigg]+    \frac{\beta_2}{m^2}\bigg[ \widehat{G}^0 {G}_i - \alpha_1
    m^2  \widehat{F}^0 {F}_i \bigg]+\\\label{Ti}&+
    \frac{1}{m^2}\frac{\beta_2^2+\gamma\beta_1}{\beta_1-\beta_2\alpha_2-\gamma\alpha_1}
    G_i\theta+\frac{1}{m}\frac{\beta_1\beta_2+\gamma(\beta_1\alpha_2-\beta_2\alpha_1)}
    {\beta_1-\beta_2\alpha_2-\gamma\alpha_1}F_i\theta=0\,,
\end{align}
where $\widehat{F}^0,\widehat{G}^{0}$ denote expressions
(\ref{Etilde}) and $\varepsilon_{12}=\varepsilon^{12}=1$.

Let us comment on the origin and meaning of these equations. The
equations (\ref{dG}) represent $i$-th component of the original
field equations (\ref{EoM}). The equation (\ref{Gauss}) is
$0$-component of the equations (\ref{EoM}), and it is understood as
Gauss law constraint of this theory (\ref{Etilde}). Relation
(\ref{dA}) is equivalent to definition (\ref{F}) of the variable
$F_i$, while (\ref{dF}) is a definition of $G_i$ (\ref{G}) being
resolved with respect to $\dot{F}$. The equations (\ref{dK}),
(\ref{T}) and (\ref{Ti}) represent the space-space, time-time, and
space-time components of Einstein's equations (\ref{ET}), with all
the time derivatives of the matter fields being expressed by means
of equations (\ref{dA}), (\ref{dF}), (\ref{dG}). The equations
(\ref{T}), (\ref{Ti}) are the constraints as they do not involve
time derivatives of the variables of the first order formulation.
These equations can be thought of as energy and momentum
constraints, given their dependence on gravitational variables. Also
notice that the matter contribution to the energy constraint
(\ref{T}) corresponds to the expression (\ref{T00-series-WEC}). The
relations (\ref{dg}) are equivalent to the definition of momentum
(\ref{pi}) and extrinsic curvature (\ref{K}) in terms of metrics.

The equivalence of these first order equations to the original ones
can be easily verified. The relations (\ref{dA}), (\ref{dF}),
(\ref{dg}) allow one to express the variables $G, F,\pi$ in terms of
$A,g$ in the form (\ref{K}), (\ref{pi}), (\ref{F}), (\ref{G}). Upon
substitution $G, F,\pi$ as functions of $A,\buildrel\ast\over{g}$
and their derivatives into the rest of equations, one arrives at the
original system (\ref{EoM}), (\ref{ET}).

Notice that the first order equations (\ref{dA})-(\ref{Ti}) cannot
be deduced from the original equations by any Legendre
transformation, because the system (\ref{EoM}), (\ref{ET}) is
non-Lagrangian with the general tensor (\ref{CT-series}) inserted in
the rhs of Einstein's equations (\ref{ET}).

In the next section, we find the Poisson bracket among the variables
$A_i,F_i,G_i, \buildrel\ast\over{g}{}_{ij}, \pi_{ij}$ such that the
equations (\ref{dA})-(\ref{Ti}) represent the first class
constrained Hamiltonian system, with $A_0, N, N^i$ being the
Lagrange multipliers at constraints $\theta, \tau, \tau_i$.

\section{Poisson brackets and Hamiltonian equations}
In previous section, we have described the first order formulation
for the system of the third order extension of Chern-Simons
(\ref{EoM}) coupled to Einstein's gravity  (\ref{ET}) through
transverse not necessarily canonical energy-momentum tensor
(\ref{CT-series}). The first order formulation includes the
evolutionary equations resolved with respect to the time derivatives
of $A_i, F_i, G_i, \buildrel\ast\over{g}{}_{ij}, \pi^{ij} $
(\ref{dA}), (\ref{dF}), (\ref{dG}), (\ref{dg}), (\ref{dK}). There
are also constraints (\ref{Gauss}), (\ref{T}), (\ref{Ti}) imposed on
the same variables. The time derivatives of $A_0$, $N$, $N^i$ are
not involved in the equations, while the variables themselves are
linearly included in the rhs of evolutionary equations (\ref{dA}),
(\ref{dF}), (\ref{dG}), (\ref{dg}), (\ref{dK}). This structure
resembles the equations of Hamiltonian constrained system. By
itself, this structure does not mean that the Poisson bracket exists
such that the rhs of evolutionary equations are the Poisson brackets
with Hamiltonian being the linear combination of constraints.
Moreover, any system of
 reparameterization invariant differential equations can be cast into
the first order normal form like that \cite{LS09}:
\begin{equation}\label{normal}
    \dot{y}^I=Z^I_\alpha(y)\lambda^\alpha \, ,
\end{equation}
\begin{equation}\label{Ty}
T_a(y)=0 \,.
\end{equation}
These equations constitute the constrained Hamiltonian system if the
Poisson bracket $\{y^I,y^J\}$ exists of the variables $y^I$ such
that
\begin{equation}\label{PBy}
Z^I_\alpha(y)\lambda^\alpha =\{y^I, T_a\lambda^a\} \equiv
\{y^I,y^J\}\partial_J T_a\lambda^a\, ,\qquad (\text{mod } T_a(y)),
\end{equation}
where $(\text{mod } T_a(y))$ means that the equality holds true up
to the terms vanishing on constraints $T$. Once the Poisson brackets
exist obeying (\ref{PBy}), the Hamiltonian is defined as the linear
combination of constraints,
\begin{equation}\label{H-t}
H(y,\lambda)= \lambda^a T_a (y)\, ,
\end{equation}
and the equations (\ref{normal}), (\ref{Ty}) read as the constrained
Hamiltonian system,
\begin{equation}\label{H-sys}
    \dot{y}=\{y\, , \, H(y,\lambda)\}\, , \qquad T(y)=0
\end{equation}
Not any system of the normal form (\ref{normal}), (\ref{Ty}) admits
the Poisson brackets obeying the conditions (\ref{PBy}). Given the
equations of motion (\ref{normal}), and constraints (\ref{Ty}), the
relations (\ref{PBy}) can be considered as equations that define the
bracket $\{y^I,y^J\}$ which endows the system (\ref{normal}),
(\ref{Ty}) with the structure of constrained Hamiltonian dynamics
(\ref{H-sys}). For general discussion of the Poisson structures
compatible with normal forms of dynamical equations see \cite{LS09}.
Below, we are seeking for the Poisson brackets such that meet the
conditions (\ref{PBy}) for the first order formulation
(\ref{dA})-(\ref{Ti}) of the original third order system
(\ref{EoM}), (\ref{ET}).

The first order formulation (\ref{dA})-(\ref{Ti}) of the original
third order equations (\ref{EoM}), (\ref{ET}) have the form
(\ref{normal}), (\ref{Ty}), with (\ref{dA})-(\ref{dG}), (\ref{dg}),
(\ref{dK}) corresponding to the evolutionary equations
(\ref{normal}). Relations (\ref{Gauss}), (\ref{T}), (\ref{Ti}) are
constraints. We chose the ansatz for the Hamiltonian in slightly
more general  form than (\ref{H-t}):
\begin{equation}\label{Htot}\begin{array}{l}\displaystyle
    H(A,g;\alpha,\beta,\gamma)=\int
    d^2x\sqrt{\buildrel\ast\over{g}}\Big\{N\tau+N^i\tau_i+k_0A_0\theta\Big\}\,,\qquad
    k_0=\frac{\beta_1^2 -\beta_1\beta_2\alpha_2 + \beta_2^2\alpha_1}{\beta_1 - \beta_2\alpha_2 -
    \gamma\alpha_1}\,,
\end{array}\end{equation}
where the constraints $\tau,\,\tau_i,\,\theta$ are defined by
relations (\ref{T}), (\ref{Ti}), (\ref{Gauss}). This ansatz includes
five independent parameters $\alpha_1, \alpha_2, \beta_1,
\beta_2,\gamma$ involved in the constraints. The parameters $\alpha$
specify the equations for the vector field (\ref{EoM}). The
parameters $\beta,\gamma$ define the matter contribution to the rhs
of Einstein's equations (\ref{ET}). These parameters are involved
into the lapse and shift constraints $\tau,\tau_i$ (\ref{T}),
(\ref{Ti}). The coefficient $k_0$ is a constant factor at the
Lagrange multiplier $A_0$. We introduce $k_0$ to conveniently
control an overall multiplier at the gradient term $\partial_iA_0$
in equations (\ref{dA}). As the constraints are defined modulo
overall non-vanishing factors, the redefinition is inessential.

We seek for the Poisson bracket assuming that the evolutionary
equations (\ref{dA}), (\ref{dF}), (\ref{dG}), (\ref{dg}), (\ref{dK})
should have the Hamiltonian form (\ref{H-sys}). This leads us to the
system of linear algebraic equations (\ref{PBy}) for unknown Poisson
brackets of the fields $A_i, F_i, G_i, \buildrel\ast\over{g}_{ij},
\pi^{ij}$:
\begin{align}\label{AH}
        &\{\,A_i\,,H\,\}_{\alpha,\beta,\gamma}=N \sqrt{\buildrel\ast\over{g}}\,\, \varepsilon_{is}\, \buildrel\ast\over{g}{\!}^{sr} F_r +
    N^k \varepsilon_{ki}\,\, \varepsilon^{sr}\, \partial_s A_r  +  \partial_i  A_0\,;\\[3mm]
        &\{\,F_i\,,H\,\}_{\alpha,\beta,\gamma}= N \sqrt{\buildrel\ast\over{g}}\,\, \varepsilon_{is}\, \buildrel\ast\over{g}{\!}^{sr} G_r +
    N^k \varepsilon_{ki}\,\,  \varepsilon^{sr}\, \partial_s F_r  +
    \partial_i \bigg( N\, \frac{1}{\sqrt{ \buildrel\ast\over{g}}}\,\, \varepsilon^{sr}\, \partial_{s}  A_{r} +  N^s  F_s \bigg)\,;\\\notag
        &\{\,G_i\,,H\,\}_{\alpha,\beta,\gamma}=- \alpha_2 m  N \sqrt{ \buildrel\ast\over{g}}\, \varepsilon_{is} \buildrel\ast\over{g}{\!}^{sr} G_r -
    \alpha_1 m^2  N \sqrt{ \buildrel\ast\over{g}}\, \varepsilon_{is}
    \buildrel\ast\over{g}{\!}^{sr} F_r - \alpha_2 m   N^k
    \varepsilon_{ki}\,\, \varepsilon^{sr}\, \partial_s F_r  -
    \alpha_1 m^2  N^k \varepsilon_{ki}\,\, \varepsilon^{sr}\, \partial_s A_r  +
    \\[3mm]\label{GH}&
    + \partial_i  \bigg( N\, \frac{1}{\sqrt{ \buildrel\ast\over{g}}}\,\, \varepsilon^{sr}\, \partial_{s} F_{r} + N^s  G_s\bigg)\,;\\
        &\{\,\buildrel\ast\over{g}_{ij}\,,H\,\}_{\alpha,\beta,\gamma}=\, \buildrel\ast\over{\nabla}_i N_j\ +\ \buildrel\ast\over{\nabla}_j N_i\ -\ 2N\,
    \frac{1}{\sqrt{\buildrel\ast\over{g}}}(\pi_{ij}-\buildrel\ast\over{g}{\!}_{ij}\pi)\,;
\end{align}
\vspace{-1.0cm}
\begin{align}\notag
    &\{\,\pi^{ij}\,,H\,\}_{\alpha,\beta,\gamma}=
    - \pi^{is} \buildrel\ast\over{\nabla}_s N^j -
    \pi^{js} \buildrel\ast\over{\nabla}_s N^i   +
    \sqrt{\buildrel\ast\over{g}} \buildrel\ast\over{\nabla}_s
    \Big(\frac{1}{\sqrt{\buildrel\ast\over{g}}} \pi^{ij} N^s\Big)
    +\\[-1mm]\notag&
    + \frac{1}{2} N \frac{1}{\sqrt{\buildrel\ast\over{g}}}
    \buildrel\ast\over{g}{}^{ij}\ ( \pi_{sr} \pi^{sr}- \pi^2 ) +
    \sqrt{\buildrel\ast\over{g}}\ \bigg[  \buildrel\ast\over{\nabla}{}^i
    \buildrel\ast\over{\nabla}{}^j  N - \buildrel\ast\over{g}{}^{ij}
    \buildrel\ast\over{\nabla}_s \buildrel\ast\over{\nabla}{}^s N \bigg]
    -   \frac{1}{2} N \sqrt{\buildrel\ast\over{g}}
    \buildrel\ast\over{g}{}^{ij}  \Lambda   -\\\notag& - \frac{1}{2}
    \frac{\beta_1}{m}N \sqrt{\buildrel\ast\over{g}}\,
    \buildrel\ast\over{g}{\!}^{ik}\buildrel\ast\over{g}{\!}^{jl} \bigg[
    \bigg( G_k F_l  + G_l F_k  - \buildrel\ast\over{g}_{kl} (
    \widehat{G}^0 \widehat{F}^0 +  \buildrel\ast\over{g}{\!}^{sr} G_s\,
    F_r ) \bigg)  + \alpha_2 m \bigg( F_k F_l  - \frac{1}{2}
    \buildrel\ast\over{g}_{kl}
    (\widehat{F}^0 \widehat{F}^0 +  \buildrel\ast\over{g}{\!}^{sr} F_s F_r) \bigg)
    \bigg] - \\[3mm]\label{piH}& - \frac{1}{2}  \frac{\beta_2}{m^2} N
    \sqrt{\buildrel\ast\over{g}}\,
    \buildrel\ast\over{g}{\!}^{ik}\buildrel\ast\over{g}{\!}^{jl} \bigg[
    \bigg( G_k G_l  - \frac{1}{2} \buildrel\ast\over{g}_{kl} (
    \widehat{G}^0 \widehat{G}^0 +  \buildrel\ast\over{g}{\!}^{sr} G_s
    G_r )\bigg) - \alpha_1 m^2 \bigg( F_k F_l - \frac{1}{2}
    \buildrel\ast\over{g}_{kl} (\widehat{F}^0 \widehat{F}^0 +
    \buildrel\ast\over{g}{\!}^{sr} F_s F_r) \bigg) \bigg]\,.
\end{align}
Given the Hamiltonian (\ref{Htot}), the solution to this system
reads
\begin{align}\label{wgpi}
    \{\, \buildrel\ast\over{g}{\!}_{ij}(\vec{x})\, ,\, \pi^{kl}(\vec{y})\, \}_{\alpha,\beta,\gamma}\ =&\
    \frac{1}{2}\, (\delta_i^k\delta_j^l+\delta_i^l\delta_j^k)\, \delta(\vec{x}-\vec{y})\,;\\[3mm]\label{wGG}
    \{\,G_i(\vec{x})\,, G_j(\vec{y})\,\}_{\alpha,\beta,\gamma\,}=&\ m^3\, \frac{\beta_1\ (\alpha_1 - \alpha_2^2) \ +\ \beta_2 \alpha_1 \alpha_2}{\beta_1^2 -\beta_1\beta_2\alpha_2 + \beta_2^2\alpha_1}\, \varepsilon_{ij}\, \delta(\vec{x}-\vec{y})\,;\\[3mm]\label{wFG}
    \{\,F_i(\vec{x})\,, G_j(\vec{y})\,\}_{\alpha,\beta,\gamma\,}=&\ m^2\, \frac{\beta_1\alpha_2 - \beta_2\alpha_1}{\beta_1^2 -\beta_1\beta_2\alpha_2 + \beta_2^2\alpha_1}\, \varepsilon_{ij}\, \delta(\vec{x}-\vec{y})\,;\\[3mm]\label{wAG}
    \{\,A_i(\vec{x})\,, G_j(\vec{y})\,\}_{\alpha,\beta,\gamma\,}=&\  \{F_i(\vec{x}), F_j(\vec{y})\}_{\alpha,\beta,\gamma}=\ m\, \frac{-\beta_1}{\beta_1^2 -\beta_1\beta_2\alpha_2 + \beta_2^2\alpha_1}\, \varepsilon_{ij}\, \delta(\vec{x}-\vec{y})\,;\\[3mm]\label{wAF}
    \{\,A_i(\vec{x})\,, F_j(\vec{y})\,\}_{\alpha,\beta,\gamma\,}=&\ \frac{\beta_2}{\beta_1^2 -\beta_1\beta_2\alpha_2 + \beta_2^2\alpha_1}\, \varepsilon_{ij}\, \delta(\vec{x}-\vec{y})\,;\\[3mm]\label{wAA}
    \{\,A_i(\vec{x})\,, A_j(\vec{y})\,\}_{\alpha,\beta,\gamma\,}=&\ \frac{1}{m}\, \frac{\gamma}{\beta_1^2 -\beta_1\beta_2\alpha_2 + \beta_2^2\alpha_1}\
    \, \varepsilon_{ij}\, \delta(\vec{x}-\vec{y})\,,\\[3mm]\notag
    \{\,\buildrel\ast\over{g}{\!}_{ij}\,,\buildrel\ast\over{g}{\!}_{kl}\,\}_{\alpha,\beta,\gamma}=&\ \{\,\pi^{ij}\,,\pi^{kl}\,\}_{\alpha,\beta,\gamma}=\
    \{\,\buildrel\ast\over{g}{\!}_{ij}\,,A_k\,\}_{\alpha,\beta,\gamma}=\
    \{\,\buildrel\ast\over{g}{\!}_{ij}\,,F_k\,\}_{\alpha,\beta,\gamma}=
    \{\,\buildrel\ast\over{g}{\!}_{ij}\,,G_k\,\}_{\alpha,\beta,\gamma}=\\[3mm]\label{zero}=
    &\ \{\,\pi^{ij}\,,A_k\,\}_{\alpha,\beta,\gamma}=\ \{\,\pi^{ij}\,,F_k\,\}_{\alpha,\beta,\gamma}=\ \{\,\pi^{ij}\,,G_k\,\}_{\alpha,\beta,\gamma}=0\,,
\end{align}
where the vectors $\vec{x}$, $\vec{y}$ label the space points, and
$\delta(\vec{x}-\vec{y})$ is the $2d$ $\delta$-function. As is seen
from the equations (\ref{wgpi})-(\ref{zero}), the Poisson bracket of
the gravity variables $\buildrel\ast\over{g}{\!}_{ij}\,,\pi^{kl}$
(\ref{wgpi}) is canonical. This happens because the Einstein's
equations without matter are Lagrangian in themselves, while the
matter contribution to the Einstein's equations (\ref{ET}), being
expressed in terms of the variables $F,G$ (\ref{FG-notation}), does
not involve derivatives of the metric. The Poisson brackets
(\ref{wGG})-(\ref{wAA}) of the vector variables $A,F,G$ involve the
parameters $\alpha,\beta,\gamma$ of the model (\ref{EoM}),
(\ref{ET}), and they are non-canonical. The parameter $\gamma$
contributes to the bracket (\ref{wAA}) between space components of
the original vector field $A_i$, while the Poisson bracket of the
physical observables $G_i,F_i,F_0$ does not depend on $\gamma$. This
is no surprise because coupling constants in the theory (\ref{EoM}),
(\ref{ET}) are $\beta_1,\beta_2$, while $\gamma$ labels equivalent
representatives of one and the same theory, so it can only
contribute to the brackets of non-gauge invariant quantities.

All the Poisson brackets are well-defined if the $00$-component
$T^{00}(\alpha,\beta)$ of the matter contribution to Einstein's
equations (\ref{ET}) is a non-degenerate quadratic form of the
fields $F,G$ (\ref{FG-notation}), see condition (\ref{b3-neq-0}). In
this way we see that the theory (\ref{EoM}), (\ref{ET}) is
Hamiltonian with the Hamilton function (\ref{Htot}) and Poisson
brackets (\ref{wgpi})-(\ref{zero}).

Given the Poisson brackets (\ref{wgpi})-(\ref{zero}), consider the
PB algebra of the constraints $\tau$ (\ref{T}), $\tau_i$ (\ref{Ti}),
$\theta$ (\ref{Gauss}). Introducing the test functions
$\mathcal{M}(x),\mathcal{M}^i(x),\mathcal{A}(x)$, we define the
functionals of constraints $T[\,\mathcal{M}\,]$,
$D[\,\mathcal{M}\,]$, $\Theta[\,\mathcal{A}\,]$:
\begin{equation}\label{T-func}
T[\,\mathcal{M}\,]=\int d^2x \sqrt{\buildrel\ast\over{g}}\
\mathcal{M}(x) \tau(x)\,,\qquad D[\,\mathcal{M}\,]=\int d^2x
\sqrt{\buildrel\ast\over{g}}\ \mathcal{M}^s(x) \tau_s (x)\,,\qquad
\Theta[\,\mathcal{A}\,]=\int d^2x \sqrt{\buildrel\ast\over{g}}\
\mathcal{A}(x) \theta(x)\,.
\end{equation}
Then, these functional have the following Poisson brackets:
\begin{align}&\notag
 \{\, T [\,\mathcal{M}\, ]\, ,  T[\,\mathcal{N}\, ]\, \} =
 D \big[\, \mathcal{N}\buildrel\ast\over{\nabla}{\!}^i \mathcal{M} -
 \mathcal{M} \buildrel\ast\over{\nabla}{\!}^i \mathcal{N}\,\big]\,,\\[4mm]\notag
&\{\, D [\, \mathcal{M}\, ]\, , T[\,\mathcal{N}\, ] \,  \}
  =     T \big[\,\mathcal{M}^s \,\partial_s \mathcal{N}\,\big] -
 \int d^2 x\, \bigg\{ k_1 \, \mathcal{N}\,  \mathcal{M}_s\, \varepsilon^{sr}  F_r\, \theta  +
 \frac{1}{2}\, k_2\, \sqrt{\buildrel\ast\over{g}}\,
\big( \,\mathcal{N} \buildrel\ast\over{\nabla}_s  \mathcal{M}^s -
\mathcal{M}^s \buildrel\ast\over{\nabla}_s \mathcal{N}\,\big)\,\theta^2 \bigg\}\,,\\[3mm]\notag
&\{\, D \big[\,\mathcal{M}\, ]\, , D [\, \mathcal{N}\, ]\, \}  = D
[\, \mathcal{M}^s  \buildrel\ast\over{\nabla}_s \mathcal{N}^i -
\mathcal{N}^s   \buildrel\ast\over{\nabla}_s \mathcal{M}^i\, \big] +
 \int d^2 x \, \bigg\{ k_1\, \sqrt{\buildrel\ast\over{g}} \,\varepsilon_{sr} \, \mathcal{M}^s \mathcal{N}^r\,
 \varepsilon^{kl} \partial_k A_l \,  \theta +
k_3 \buildrel\ast\over{g} \varepsilon_{sr}\, \mathcal{M}^s
\mathcal{N}^r\,  \theta^2 \,\bigg\}\,,
\\[3mm]\notag
&\{\, T [\,\mathcal{M}\, ]\,, \Theta [\mathcal{A}\, ]\,  \}\  = 0\,,
\qquad \{\,D [\, \mathcal{M}\, ]\,, \Theta [\mathcal{A}\, ]\, \}\ =
0\,, \qquad \{\, \Theta [\mathcal{M}\, ]\,, \Theta [\mathcal{A}\,
]\, \}\ =
0\,,\\[3mm]
\label{PBTT}&
    k_1 = \frac{\beta_1^2 -\beta_1\beta_2\alpha_2 + \beta_2^2\alpha_1}{\beta_1 - \beta_2\alpha_2 -
    \gamma\alpha_1}\,,
\quad k_2 = \frac{1}{m^2}\frac{\gamma^2 (\beta_1\alpha_2
-\beta_2\alpha_1) + 2\gamma\beta_1\beta_2+\beta_2^3}{(\beta_1 -
\beta_2\alpha_2 - \gamma\alpha_1)^2}\,, \quad k_3 =
\frac{1}{m}\frac{\gamma (\beta_1^2 -\beta_1\beta_2\alpha_2 +
\beta_2^2\alpha_1)}{(\beta_1 - \beta_2\alpha_2 -
\gamma\alpha_1)^2}\,.
\end{align}
As is seen, the  PBs of constraints $\tau,\tau_i$  correspond to the
algebra of lapse and shift transformations modulo the Gauss law
constraint $\theta$. The Gauss law constraint $\theta$ has
identically vanishing Poisson brackets with all the constraints.
These PB relations indicate that the constraints $\tau, \tau_i$
generate the lapse and shift transformations for all the fields in
theory, while the Gauss law constraint $\theta$ (\ref{Gauss})
generates the gradient gauge transformations for components of
original vector field $A_i$ as it does in free theory without
matter.

Let us provide explicit expressions for Poisson brackets of the
fields $\buildrel\ast\over{g}_{ij},\pi^{ij},G_i,F_i,A_i$ with the
constraints
$T[\,\mathcal{M}\,],D[\,\mathcal{M}\,],\Theta[\,\mathcal{A}\,]$
(\ref{T-func}):
\begin{align}\label{tA}&
        \{\, A_i\,, T [\,\mathcal{M}\,]\, \}_{\alpha,\beta,\gamma}  =
    \mathcal{M} \sqrt{\buildrel\ast\over{g}} \, \varepsilon_{is}
    \buildrel\ast\over{g}{\!}^{sr} F_r - \frac{ \gamma^2(\beta_1\alpha_2
    - \beta_2\alpha_1)+   2\gamma\beta_1\beta_2 +\beta_2^3 }{(\beta_1^2
    -\beta_1\beta_2\alpha_2 + \beta_2^2\alpha_1)(\beta_1 -
    \beta_2\alpha_2 - \gamma\alpha_1)} \frac{1}{m^2} \partial_i \left(
    \mathcal{M}\,\theta   \right) \,;\\[4mm]\label{tF}
       & \{\,  F_i\,, T[\,\mathcal{M}\,] \,   \}_{\alpha,\beta,\gamma}   =
    \mathcal{M} \sqrt{\buildrel\ast\over{g}}\, \varepsilon_{is} \buildrel\ast\over{g}{\!}^{sr}  {G}_r  +
    \partial_i\Big(  \mathcal{M}\frac{1}{ \sqrt{\buildrel\ast\over{g}}} \, \varepsilon^{sr} \partial_s A_r     +
    \frac{\gamma }{\beta_1 - \beta_2\alpha_2 - \gamma\alpha_1}  \frac{1}{m}   \mathcal{M}\,\theta
    \Big)\,;\\[2mm]\notag
        &\{\, G_i\,, T[\,\mathcal{M}\,]\,  \}_{\alpha,\beta,\gamma}    =
    \alpha_2  m    \mathcal{M} \sqrt{\buildrel\ast\over{g}}\,
    \varepsilon_{si} \buildrel\ast\over{g}{\!}^{sr}  {G}_r  + \alpha_1
    m^2 \mathcal{M}  \sqrt{\buildrel\ast\over{g}}\, \varepsilon_{si}
    \buildrel\ast\over{g}{\!}^{sr}  {F}_r +\\[3mm]\label{tG}& +
    \partial_i \Big(  \mathcal{M} \frac{1}{\sqrt{\buildrel\ast\over{g}}} \, \varepsilon^{sr} \partial_s F_r   +
    \frac{\beta_2}{\beta_1 - \beta_2\alpha_2 - \gamma\alpha_1}
    \mathcal{M} \, \theta   \Big)\,;\\\label{tg}
        &\{\, \buildrel\ast\over{g}_{ij}\,, T[\,\mathcal{M}\,]\,   \}_{\alpha,\beta,\gamma}  =
    - 2 \mathcal{M} \frac{1}{\sqrt{\buildrel\ast\over{g}}} ( \pi_{ij} - \buildrel\ast\over{g}{}_{ij}
    \pi)\,;\\[1mm]\notag
        &\{\, \pi^{ij}\, , T[\,\mathcal{M}\,]\,   \}_{\alpha,\beta,\gamma}  =\frac{1}{2}\, \mathcal{M} \frac{1}{\sqrt{\buildrel\ast\over{g}}} \buildrel\ast\over{g}{\!}^{ij}
    \left( \pi_{sr}\pi^{sr} -  \pi^2 \right)
    + \sqrt{\buildrel\ast\over{g}} \left[\buildrel\ast\over{\nabla}{\!}^i \buildrel\ast\over{\nabla}{\!}^j \mathcal{M }-
    \buildrel\ast\over{g}{\!}^{ij}  \buildrel\ast\over{\nabla}_s \buildrel\ast\over{\nabla}{\!}^s \mathcal{M} \right]
    - \frac{1}{2}\ \mathcal{M}\sqrt{\buildrel\ast\over{g}} \buildrel\ast\over{g}{\!}^{ij}  \Lambda
    - \\[3mm]\notag& - \frac{1}{2}\frac{\beta_1}{m} \mathcal{M} \sqrt{\buildrel\ast\over{g}}\,
    \buildrel\ast\over{g}{\!}^{ik}\buildrel\ast\over{g}{\!}^{jl} \bigg[
    \bigg( G_k F_l  + G_l F_k  - \buildrel\ast\over{g}_{kl} (
    \widehat{G}^0 \widehat{F}^0 +  \buildrel\ast\over{g}{\!}^{sr} F_s
    G_r ) \bigg)  + \alpha_2 m \bigg( F_k F_l  - \frac{1}{2}
    \buildrel\ast\over{g}_{kl}
    (\widehat{F}^0 \widehat{F}^0 +  \buildrel\ast\over{g}{\!}^{sr} F_s F_r) \bigg)
    \bigg] -\\[3mm]\notag&  - \frac{1}{2}\frac{\beta_2}{m^2}\mathcal{M}
    \sqrt{\buildrel\ast\over{g}}\,
    \buildrel\ast\over{g}{\!}^{ik}\buildrel\ast\over{g}{\!}^{jl} \bigg[
    \bigg( G_k G_l - \frac{1}{2} \buildrel\ast\over{g}_{kl} (
    \widehat{G}^0 \widehat{G}^0 +  \buildrel\ast\over{g}{\!}^{sr} G_s
    G_r )\bigg) - \alpha_1 m^2 \bigg( F_k F_l - \frac{1}{2}
    \buildrel\ast\over{g}_{kl} (\widehat{F}^0 \widehat{F}^0 +
    \buildrel\ast\over{g}{\!}^{sr} F_s F_r) \bigg)
    \bigg]+\\[3mm]\label{tpi}&
    + \frac{1}{2}\,  \mathcal{M} \buildrel\ast\over{g}{}^{ij} \left(
    \frac{\beta_1\beta_2 + \gamma (\beta_1\alpha_2 -
    \beta_2\alpha_1)}{\beta_1 - \beta_2\alpha_2 - \gamma\alpha_1}
    \frac{1}{m} \, \varepsilon^{sr} \partial_s A_r  + \frac{\beta_2^2 +
    \gamma\ \beta_1}{\beta_1 - \beta_2\alpha_2 - \gamma\alpha_1}
    \frac{1}{m^2} \, \varepsilon^{sr} \partial_s F_r   \right) \theta\,,
    \\[3mm]\label{tiA}
        &\{\,  A_i\,, D[\, \mathcal{M}\, ]\,   \}_{\alpha,\beta,\gamma}   =
    \mathcal{M}^s\ \partial_s   A_i  + A_s \partial_i \mathcal{M}^s -
    \partial_i \big( \mathcal{M}^s {A}_s   \big) +
    \frac{\gamma}{\beta_1 - \beta_2\alpha_2 - \gamma\alpha_1}
    \frac{1}{m} \mathcal{M}^s \sqrt{\buildrel\ast\over{g}} \, \varepsilon_{si} \,
    \theta\,;\\[3mm]\label{tiF}
        &\{\,  F_i\,, D [\,\mathcal{M}\, ]\,   \}_{\alpha,\beta,\gamma}   =
    \mathcal{M}^s \partial_s   F_i  + F_s \partial_i \mathcal{M}^s  +
    \frac{\beta_2}{\beta_1 - \beta_2\alpha_2 - \gamma\alpha_1} \mathcal{M}^s
    \sqrt{\buildrel\ast\over{g}} \, \varepsilon_{si} \,
    \theta\,;\\[3mm]\label{tiG}
        &\{\,  G_i\, , D[\, \mathcal{M}\, ] \,   \}_{\alpha,\beta,\gamma}   =
    \mathcal{M}^s \partial_s   {G}_i + G_s\ \partial_i \mathcal{M}^s -
    \frac{\beta_1}{\beta_1 - \beta_2\alpha_2 - \gamma\alpha_1}\ m\
    \mathcal{M}^s \sqrt{\buildrel\ast\over{g}} \, \varepsilon_{si} \,
    \theta\,;\\[3mm]\label{tig}
        &\{\,  \buildrel\ast\over{g}_{ij} , D[\, \mathcal{M}\, ] \,\}_{\alpha,\beta,\gamma}  =
    \mathcal{M }^s \partial_s g_{ij} +
    g_{is} \partial_j\mathcal{M }^s + g_{s j } \partial_i\mathcal{M
    }^s\,;\\[4mm]\label{tipi}
        &\{\,  \pi^{ij}\,, D[\, \mathcal{M}\, ]\,    \}_{\alpha,\beta,\gamma}   = - \pi^{is}
    \partial_s \mathcal{M}^j -
    \pi^{sj} \partial_s  \mathcal{M}^i +
    \partial_s \big(\mathcal{M}^s  \pi^{ij}\big)\,;\\[3mm]\label{TTA}
        &\{\,A_i\,,\Theta[\,\mathcal{A}\,]\,\}_{\alpha,\beta,\gamma} =\frac{\beta_1-\beta_2\alpha_2-\gamma\alpha_1}{\beta_1^2-\beta_1\beta_2\alpha_2+\beta_2^2\alpha_1}
        \partial_i\mathcal{A}\,,\\[3mm]\label{TTf}&
    \{\,F_i\,,\Theta[\,\mathcal{A}\,]\,\}_{\alpha,\beta,\gamma} =
    \{\,G_i\,,\Theta[\,\mathcal{A}\,]\,\}_{\alpha,\beta,\gamma}=
    \{\,\buildrel\ast\over{g}{\!}_{ij}\,,\Theta[\,\mathcal{A}\,]\,\}_{\alpha,\beta,\gamma}=
    \{\,\pi^{ij}\,,\Theta[\,\mathcal{A}\,]\,\}_{\alpha,\beta,\gamma}=0\,,
\end{align}
where $A^0$ denotes the time component of $3d$ vector $A^\alpha$,
and $\widehat{F}^0,\widehat{G}^0$ are introduced in (\ref{Etilde}).
On the shell of Gauss law constraint (\ref{Gauss}) and evolutionary
equations of motion (\ref{dA}), (\ref{dF}), (\ref{dG}), (\ref{dg}),
(\ref{dK}), the Poisson brackets (\ref{tF})-(\ref{tpi}),
(\ref{tiF})-(\ref{tipi}) correspond to the transformation law for
the quantities $F,G,\buildrel\ast\over{g},\pi$ under the action of
infinitesimal diffeomorphisms. The constraint functional
$T[\,\mathcal{M}\,]$ generates shifts along the timelike vector
$\xi=(\mathcal{M}/N,-\mathcal{M}N^i/N)$, while $D[\,\mathcal{M}\,]$
acts by space shifts, with $\zeta=(0,\mathcal{M}^i)$ being the shift
vector. These geometric inerpretations suggest that the constraints
$\tau$ and $\tau_i$ generate the lapse and shift transformations for
each tensor quantity $O(F,G,\buildrel\ast\over{g},\pi)$,
\begin{align}\label{All-T1}
    &\{\,O(F,G,\buildrel\ast\over{g},\pi )\,, T[\,\mathcal{M}\,]\,\}_{\alpha,\beta,\gamma}= L_\xi O(F,G,\buildrel\ast\over{g},\pi
    )\,,\qquad \xi=(\mathcal{M}/N,-\mathcal{M}N^i/N)\,, \\[3mm]\label{All-T}
    &\{\,O(F,G,\buildrel\ast\over{g},\pi )\,,D[\,\mathcal{M}\,]\,\}_{\alpha,\beta,\gamma}= L_\zeta
    O(F,G,\buildrel\ast\over{g},\pi)\,,\qquad
    \zeta=(0,\mathcal{M}^i)\,.
\end{align}
The  time derivatives are defined in the rhs of equations
(\ref{All-T1}) by the equations of motion (\ref{dA})-(\ref{dK}).
Relation (\ref{All-T}) holds true modulo Gauss law constraint
(\ref{Gauss}). The brackets of the field $A$ with
$T[\,\mathcal{M}\,], D[\,\mathcal{M}\,]$ represent the lapse and
shift transformations with certain admixture of gradient gauge
transformations for the vector potential\footnote{ The relation
(\ref{A-T}) is understood in the same sense as (\ref{All-T1}),
(\ref{All-T}), i.e. with the time derivatives defined by equations
of motion, and modulo Gauss constraint.},
\begin{equation}\label{A-T}
    \{\,A_i\,,T[\,\mathcal{M}\,]\,\}_{\alpha,\beta,\gamma}=(L_\xi A)_i-\partial_i(\mathcal{M}NA^0)\,, \qquad
    \{\,A_i\,,D[\,\mathcal{M}\,]\,\}_{\alpha,\beta,\gamma}=(L_\zeta
    A)_i-\partial_i(\mathcal{M}^sA_s)\,,\qquad i=1,2.
\end{equation}
The admixture of gauge transformation seems to be admissible because
the vector potential is not a gauge invariant quantity. It is not
even 1-form, being rather 1-gerbe. Because of that, it seems natural
that the lapse and shift transformations for $A$ may involve an
admixture of the gauge transformation adding the exact 1-form to the
gerbe. The gauge invariant physical observables $F$ and $G$
transform under the lapse and shift in the usual way, as one-forms.
The latter fact identifies the constraints $\tau$, $\tau_i$ with the
lapse and shift Hamiltonian generators.

In constrained Hamiltonian formalism, the strongly conserved energy
of the model is usually identified with the constraint that
generates lapse transformations. In the model (\ref{EoM}),
(\ref{ET}), the matter contribution to the lapse constraint $\tau$
(\ref{T}) is given by the $00$-component
$T^{00}(\alpha,\beta,\gamma)$ (\ref{T00-series-WEC}) of the on-shell
covariantly transverse tensor $T^{\mu\nu}(\alpha,\beta,\gamma)$
(\ref{CT-series}). Under the relations
(\ref{Stability-condidtions}), the matter contribution to the lapse
constraint is bounded. This fact provides another evidence of
stability of the model.

\section{Concluding remarks}
In this paper we focus at three aspects of the third order extension
of Chern-Simons. At first, we introduce gravity into the field
equations (\ref{EoM}) in a minimal way. Then we notice that the
theory admits a two-parameter series of on-shell covariantly
transverse tensors (\ref{CT-series}). This leaves some freedom in
consistent inclusion of the matter into the rhs of Einstein's
equations (\ref{ET}) because  any of the transverse tensors fits
well to this role. Second, we see that some of the admissible
couplings with gravity meet the weak energy condition, so they are
stable, while the interaction through the canonical energy-momentum
breaks stability. The third point is that the inclusion of
interaction with gravity through the non-canonical energy-momentum,
being a non-Lagrangian coupling, still admits constrained
Hamiltonian formulation of the corresponding first order equations.
So, the theory admits quantization while the interaction is not
necessarily Lagrangian.

Let us mention in the end, that even though Einstein's gravity does
not have its own local degrees of freedom in $3d$, inclusion of
interaction with matter subject to higher derivative equations is a
nontrivial issue, especially from the viewpoint of maintaining
stability. Even if the matter dynamics is stable in the flat space
due to conservation of certain tensor which differs from the
canonical energy-momentum, it is unclear a priori why and how the
stability can persist in the non-flat Einstein's space that
corresponds to the non-trivial energy-momentum in the rhs. This work
suggests the pattern of construction that can answer to this
question. The construction, in fact, is not too sensitive to the
dynamical content of gravity. One can expect that the similar
pattern should work in higher dimensions, where Einstein's gravity
has its own local degrees of freedom. In $d>3$, however, the
explicit construction of consistent and stable coupling of gravity
with higher derivative matter can become more complicated in certain
respect. The construction of transverse tensors meeting the WEC for
higher derivative field equations seems basically following the same
pattern in $d>3$ as in $d=3$. It is the non-canonical construction
of constrained Hamiltonian formalism which may seem a more complex
issue in $d>3$.

 \vspace{0.2cm}
 \noindent
{\bf Acknowledgments.} We thank P.O.~Kazinski and A.A.~Sharapov for
discussions. The work is partially supported by the RFBR grant
16-02-00284 and by Tomsk State University Competitiveness
Improvement Program. SLL acknowledges support from the project
3.5204.2017/6.7 of Russian Ministry of Science and Education.

\end{document}